\documentclass{article}

\usepackage{arxiv}

\usepackage[utf8]{inputenc} 
\usepackage[T1]{fontenc}    
\usepackage{hyperref}       
\usepackage{url}            
\usepackage{booktabs}       
\usepackage{amsfonts}       
\usepackage{nicefrac}       
\usepackage{microtype}      
\usepackage{lipsum}
\usepackage{graphicx}
\graphicspath{ {./images/} }

\title{Mapping and maneuvering long-term natural orbits around Titania, a satellite of Uranus.}

\author{
 Silvia Giuliatti Winter \\
  Grupo de Dinâmica Orbital e Planetologia (GDOP)\\
  São Paulo State University (UNESP)\\
 Guaratinguetá, 12516-410, SP, Brazil \\
  \texttt{giuliatti.winter@unesp.br} \\
   \And
  Jadilene Xavier \\
  Grupo de Dinâmica Orbital e Planetologia (GDOP)\\
  São Paulo State University (UNESP)\\
 Guaratinguetá, 12516-410, SP, Brazil \\
  \texttt{jadilene.rodrigues@unesp.br} \\
  \And
Antônio Bertachini Prado \\
  Postgraduate Division \\
 National Institute for Space Research (INPE)\\
São José dos Campos 12227-010, SP, Brazil \\
  \texttt{antonio.prado@inpe.br} \\
 \And
Andre Amarante \\
 Grupo de Dinâmica Orbital e Planetologia (GDOP)\\
  São Paulo State University (UNESP)\\
 Guaratinguetá, 12516-410, SP, Brazil \\
  \texttt{andre.amarante@unesp.br} \\
}

\begin{document}
\maketitle
\begin{abstract}
In this work, we present the results of a set of numerical simulations carried out to obtain long-duration orbits for a probe around Titania, Uranus' largest satellite. We also propose orbital maneuvers to extend the lifetime of some orbits. Titania's $J_2$ and $C_{22}$ gravitational coefficients and Uranus' gravitational perturbation are considered. The analysis of lifetime sensitivity due to possible errors in $J_2$ and $C_{22}$ values is investigated using multiple regression models. Simulations were performed for eccentricity equal 10-4, and lifetime maps were constructed. The results show that low-altitude orbits have longer lifetimes due to the balance between the disturbance of Uranus and the gravitational coefficients of Titania. The results also show that non-zero values of periapsis longitude ($\omega$) and ascending node longitude ($\Omega$) are essential to increase lifespan. Furthermore, the results indicate that the most economical maneuver occurs for final orbits of radius equal to 1050 km, this is observed for all inclination values.

\end{abstract}


\section{Introduction}
 Recent works have presented important studies on orbits around some natural satellites. Some of these satellites have important characteristics capable of explaining the origin of their system and, consequently, of the solar system. In this way, sending a probe to study regions close to these moons is necessary. The objective of these works is to propose better scenarios to keep a probe in orbit around these satellites for as long as possible.
In \cite{jady1} tthe orbital duration of a probe in a low-altitude orbit around Titania is investigated. The authors consider the flattening and ellipticity of Titania through the coefficients $J_2$ and $C_{22}$, in addition to the gravitational attraction of Uranus. The authors find orbits with a lifetime of up to 1000 days and show that these orbits are found at lower altitudes and with an eccentricity value equal to $10^{-3}$. In addition, through a multiple regression model, the authors investigate how possible errors in the gravitational coefficients can affect the lifetime of the probe and conclude that the orbit with $e=10^{-3}$ is the most sensitive to these errors. 
In the work of \cite{thamis2} a satellite of Jupiter, Io, is analyzed. Long-term orbits are investigated considering perturbations caused by the gravitational coefficient $J_2$ and the gravitational attraction of Jupiter. Through a set of numerical simulations, the authors show that the orbits are strongly disturbed by Jupiter and that the effects due of the flattening of Io reduce the effects of the third body in some cases. In this work, the authors highlight the importance of good choices in the values of the pericenter argument and the ascending node longitude, but they emphasize the care regarding the sensitivity of these values since any small variation in these values can cause the orbital duration to decrease. Furthermore, the authors show that as the initial orbital inclination is reduced, the lifetime lasts longer, as well as for less eccentric orbits.
    The dynamics of natural orbits around Saturn's natural satellite Titan is studied in the work of \cite{lucas3}. For this study, the effects due to the gravitational attraction of Saturn and the effects of the gravitational coefficient $J_2$ were considered. In addition, the effects of atmospheric drag were also considered in this study. Through life maps, it was possible to observe that, for all cases of initial eccentricities used, specific values of inclination and semimajor axis can generate long orbits with a lifetime of up to 20 years. In addition, it was observed that, in some regions delimited in the work, the atmospheric drag present on Titan acted, in most cases, contributing to the reduction of the orbit's lifetime, making the collision happen earlier. In order to return the probe to its original orbit and thereby increase its lifetime, the authors also propose low-fuel-cost orbital maneuvers for initial polar orbits.
       In this work, we present a study of a low-altitude, near-circular polar orbit for a probe around Titania, the largest natural satellite of Uranus. For this study, we considered the moon's $J_2$ and $C_{22}$ gravitational coefficients, as well as the gravitational perturbation due to Uranus. In order to avoid the spacecraft's collision with the surface of Titania, we propose orbital maneuvers in order to make the life time be extended.

\section{2. Mathematical model} 
     The system addressed in this work consists of a central body (Titania), a space probe, and a perturber (Uranus). We assume that the perturber body is in a Keplerian orbit around Titania with a radius of $25.362 \times 10^3$ km and mass $8.68 \times 10^{25}$ kg  \url{https://ssd.jpl.nasa.gov/}. To make an accurate investigation of the problem, we consider the main gravity coefficients ($J_2$ and $C_{22}$) of Titania, regarding the oblateness and ellipticity of the body. Titania has a mass of $35.27 \times 10^{20}$ kg and a radius of 788.9 km and the other parameters are shown in Table \ref{tab:1}. The equations of motion are described according to \cite{scheeres4}:

\begin{equation}  
\ddot{\vec{r}}= - \frac{G(M_T+m)\vec{r}}{\vec{r}^3} + 
 GM_U \left(\frac{\vec{r}_U-\vec{r}}{\mid \vec{r}_U-\vec{r} \mid ^3}-\frac{\vec{r}_U}{\vec{r}_U^3}\right) + \vec{P}_T 
\label{eq:1}
\end{equation}
The gravitational coefficients and some physical parameters of the Uranus-Titania system are presented in Table \ref{tab:1}.

\begin{table}[ht]
\centering
  \caption{Parameters of Titania with respect to Uranus.}
 \label{tab:1}
 \begin{tabular}{cc}
 \hline
   \textbf{Parameter}         & \textbf{Value} \\
  \hline
  Semi-major axis (km)    &  $435.8 \times 10^{3}$     \\
  Eccentricity            &  $1.18 \times 10^{-3}$   \\  
  Inclination     ($^\circ$)   &  $10^{-1}$                    \\
  Argument of periapsis ($^\circ$)&  $1.64 \times 10^{2}$  \\
  Longitude of ascending node ($^\circ$) & $1.67 \times 10^{2}$    \\
  Mean anomaly  ($^\circ$)     & $2.05 \times 10^{2}$       \\

Gravitational Coefficient ($J_2$) & $1.13 \times 10^{-4}$ \\ 

Gravitational Coefficient ($C_{22}$) & $3.38 \times 10^{-5}$ \\
 \hline
\end{tabular}\\
\hspace{-0.6cm}JPL.Website: \url{https://ssd.jpl.nasa.gov/}. 
 \end{table}

\section{Analyzing the Results}
     The next figures show the results of simulations made for a low-altitude orbit around a moon. Our moon Titania is the central body and Uranus acts as the third body in an elliptical orbit.. In order to analyze the influence of perturbations caused by the gravitational attraction of Uranus and Titania's $J_2$ and $C_{22}$ on the probe's lifetime, the results presented will be: (i) Considering only the perturbation of the third body. (ii) Considering the third body perturbation and Titania's gravitational coefficients.
We analyzed an orbit with eccentricity equal to $10^{-4}$ and then constructed tilt lifetime maps versus a semi-major axis with an inclination of 75 to 90 degrees and a semi-major axis of 810 to 2500 km. For the argument of the pericenter and the longitude of the ascending node, we assign a value of zero. The lifetime is represented by the color bar and the maximum orbital duration time is considered when a collision happens.

\begin{figure}[!h]
\centering
    \includegraphics[width = 6.4cm]{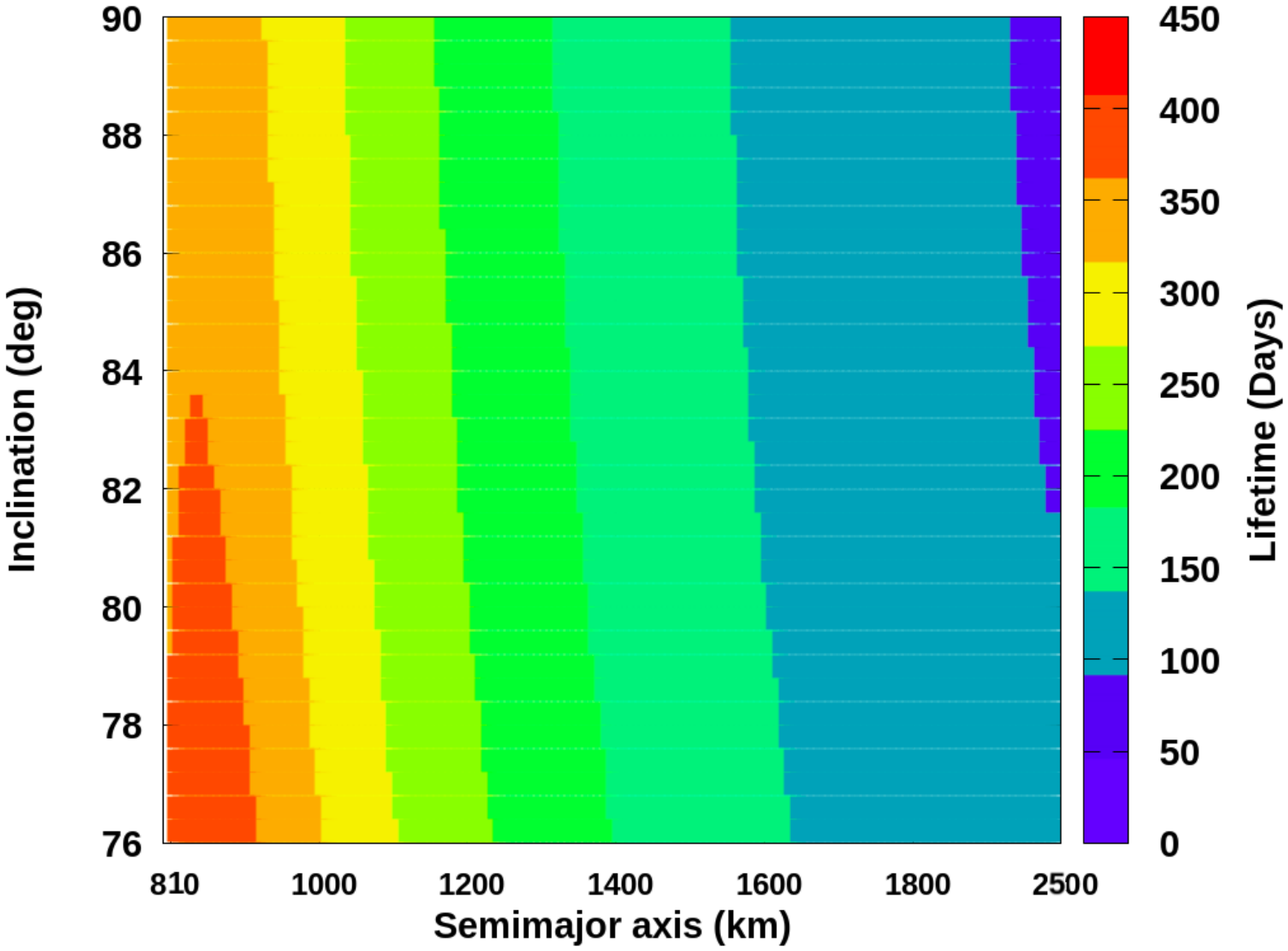} \includegraphics[width = 6.4cm]{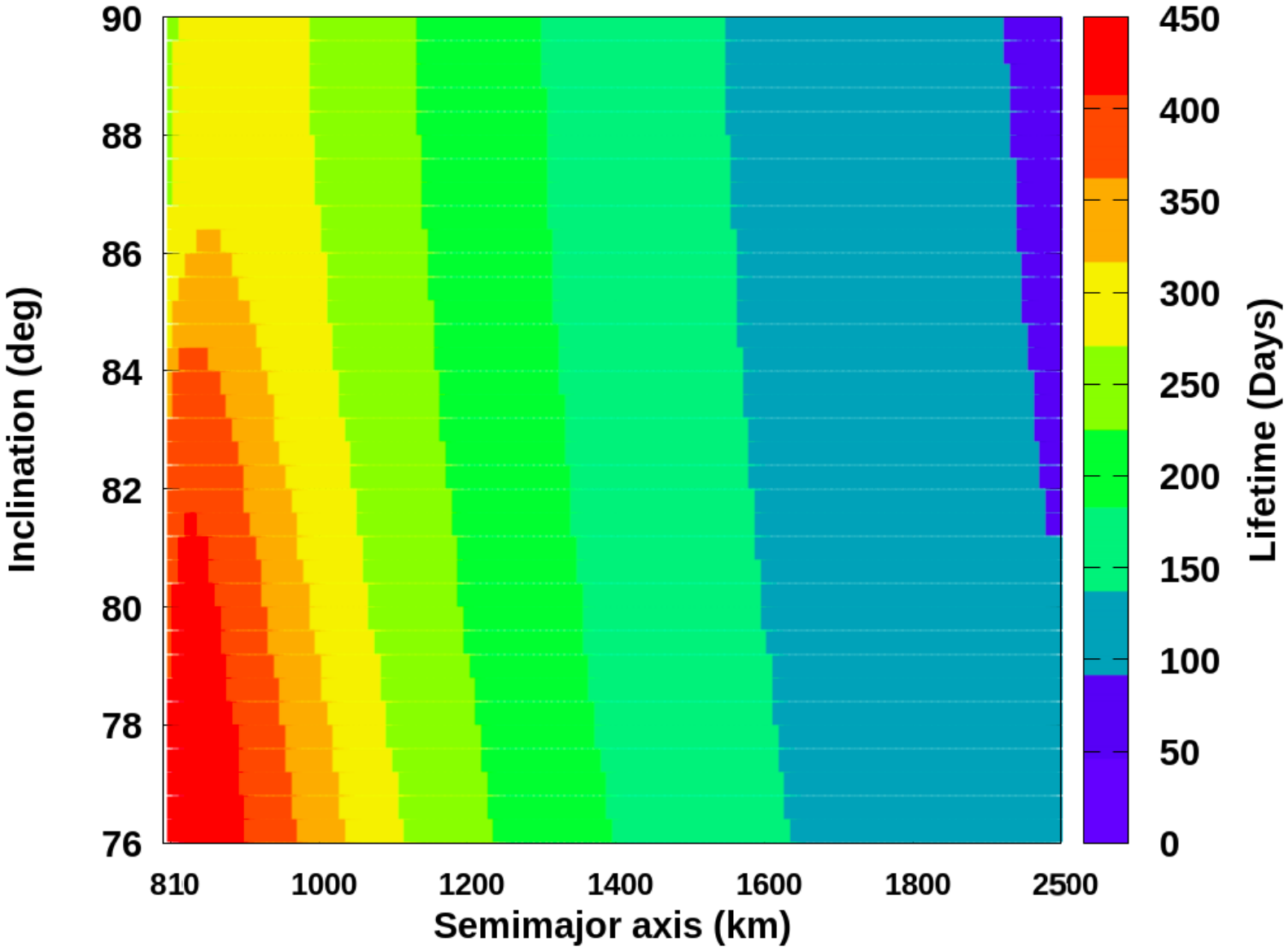}
    \caption{Diagram of a versus I for $e=10^{-4}$: a) considering only the effects from Uranus; b) including the third-body and the effects of $J_2$ and $C_{22}$ of Titania. Initial values are $a = 810-1200$ km, $I = 75-90^\circ$, $\omega = \Omega =0^\circ$.}
    \label{fig:1}
\end{figure}

 Figure \ref{fig:1}a shows the numerical simulations where the gravitational effects of Uranus are considered. Orbits with a lifetime close to these values appear only on a small island, for all inclinations, and a semimajor axis closer to 1400 km. Figure 1b, including the gravity coefficients of Titania, presents a more extensive region with lifetimes of 350–500 days. This region is located very close to the surface of Titania ($a = 810-1000$ km), where the terms $J_2$ and $C_{22}$ cancel the effects of the third body.
      In the regions where the probe remains around Titania for longer times, there is a balance between the perturbation caused by the third body and the perturbation due to the gravitational coefficients of Titania, causing the probe’s lifetime to be extended. This balance is described in previous work as a “protection mechanism” responsible for softening the effects caused by the third body perturbation on the variation of the eccentricity.
       We analyzed the regions with longer lifetimes, as shown in Figure \ref{fig:1}. Within these regions, we chose the best values of a and I and built maps as a function of $\omega$ and $\Omega$ for $e=10^{-4}$. For this analysis, we consider the perturbation due to the third body and also the gravity coefficients of Titania.

       \begin{figure}[!h]
\centering
    \includegraphics[width = 7.4cm]{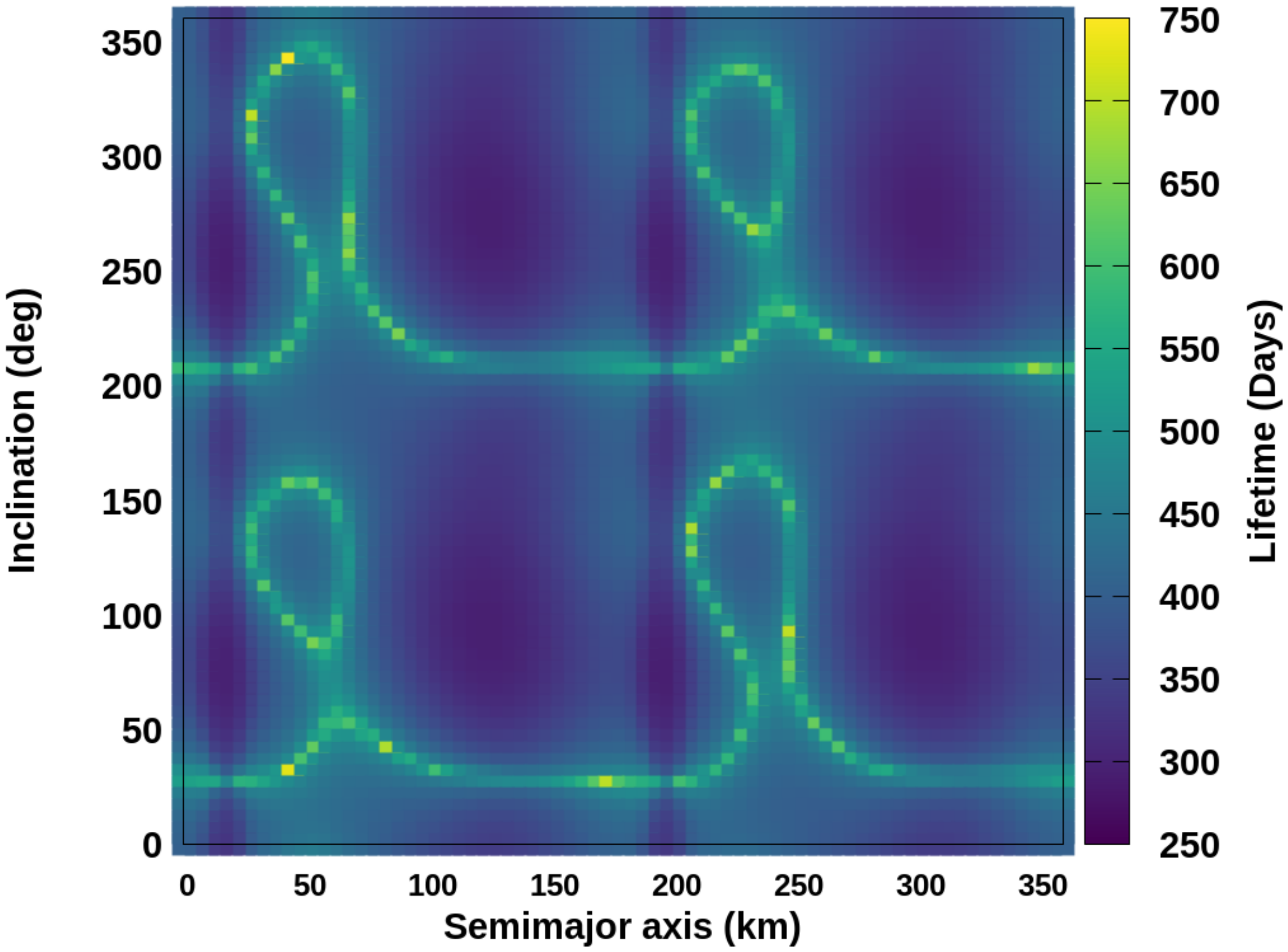} 
    \caption{Diagram of $\omega \times \Omega$ for $e=10^{-4}$ considering the effects of third-body and $J_2$ and $C_{22}$ of Titania: b) $a=930$ km, $I=80^\circ$ , $\omega = 0-360^\circ$ , $\Omega=0-360^\circ$.}
    \label{fig:2}
\end{figure}

An initial condition (a point) with values of $\omega=150^\circ$ and $\Omega=50^\circ$ is highlighted (black circle) in Figure 2. The values of these angles were added in the new simulations for the initial conditions of Figure \ref{fig:2}, in order to investigate how they would affect the lifetime of the probe. The new investigation was carried out by considering the complete system. The use of the non-zero values for these angles in $e=10^{-4}$ increases by 100\% of the lifetime of the probe.

\subsection{Analyzing the Gravitational Coefficients of Titania}

       To analyze how possible errors in the values of Titania's gravitational coefficients can affect the life of the probe. Through multiple regression models, we were able to find a relationship as a function of the coefficients $J_2$ and $C_{22}$ capable of predicting the error in the probe lifetime. We investigated a range of -10\% to 10\% for $J_2$ and $C_{22}$. We used a regression model with an accuracy coefficient of $R^2=0.85$ and the relationship found is shown in equation \ref{eq:R} To find the observed function, we consider only the case in which the lifetime variation is significant. The initial conditions for the regression were taken from the analyzed percentages for $e=10^{-3}$, as it presents the greatest increase in lifetime.

\begin{equation}
    Y=854.6+1844.6J_2 - 684J_2^2-3259.1C_{22} + 2115.5C_{22}^2-375.9 J_2C_{22}
\label{eq:R}
\end{equation}

Where Y is the probe lifetime, $J_2$ and $C_{22}$ are Titania's gravitational coefficients plus possible errors. 

\section{Orbital maneuvers}

It is not interesting that a probe around a natural satellite collides with the surface of the moon after a few days of observation. In this way, we propose bi-impulsive orbital maneuvers made from an initial elliptical orbit to a final circular orbit, both coplanar. The radii of the intended final orbits range from 1000 to 2000 km and the three values of inclinations analyzed are $70^\circ$, $80^\circ$ and $90^\circ$. The maneuver was performed days before the collision and the total velocity increment ($\Delta V$) is given by:

 \begin{equation}
 V_{ap}= \sqrt{\frac{2\mu}{a(1+e)}-\frac{\mu}{a}}
\end{equation}

    \begin{equation}
V_{p1}=\sqrt{\frac{2\mu}{a(1+e)}-\frac{2\mu}{a(1+e)+r_{cir}}}
    \label{eq:1.1} 
\end{equation}

\begin{equation}
 \Delta V_1=V_{p1}-V_{ap}
   \label{eq:2.1} 
\end{equation}

\begin{equation}
 V_{2}=\sqrt{\frac{2\mu}{r_{cir}}-\frac{2\mu}{a(1+e)+r_{cir}}}
\end{equation}
 
\begin{equation}
   V_{cir}=\sqrt{\frac{\mu}{r_{cir}}}
   \label{eq:4.1} 
\end{equation}

\begin{equation}
  \Delta V_2=V_{cir}-V_2
   \label{eq:5.1} 
\end{equation}

\begin{equation}
 \Delta V=\mid \Delta V_1 \mid + \mid \Delta V_2 \mid
   \label{eq:6.1} 
\end{equation}

Where $V_{ap}$ is the velocity at the apocenter of the initial orbit, $V_{p1}$ the velocity at point 1 of the transfer orbit. $V_{cir}$ the final speed of the circular orbit, $V_2$ is the speed of the second pulse at point 2. 

\begin{figure}[!h]
    \centering
    \includegraphics[width = 8.4cm]{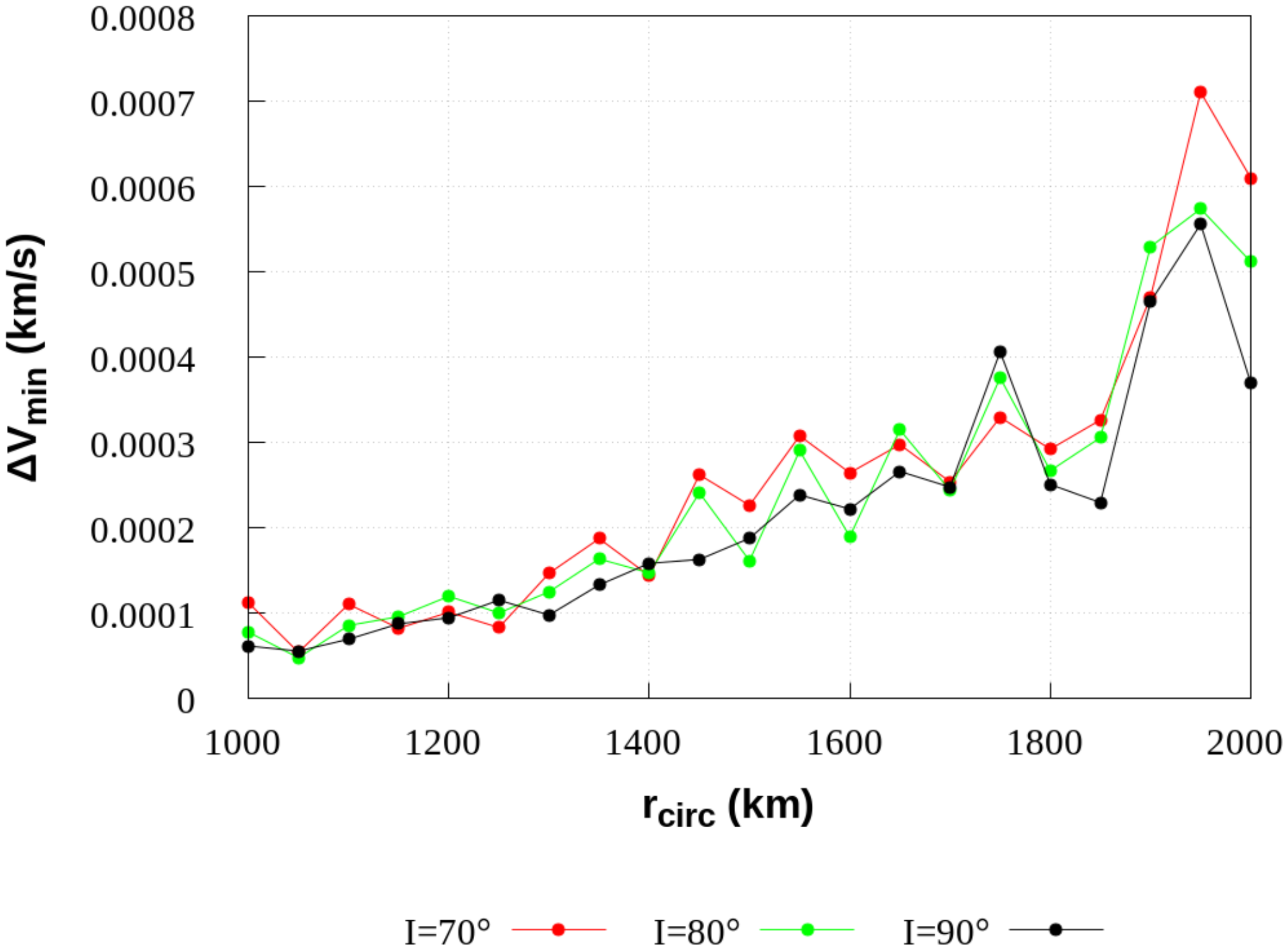}
    \caption{Minimum $\Delta V$ as a function of the circular radius of the desired final orbit. Analyzing three initial values of inclination $70^\circ$, $80^\circ$ and $90^\circ$, $a_i = 1000-2000$ km, $e=0$, $\omega=\Omega=0^\circ$. Radius of intended final orbits $1000-2000$ km.
}
    \label{fig:4}
\end{figure}

The smallest value of $\Delta V$ found was approximately $5.4 \times 10^{-5}$ km/s for a final orbit with a radius of 1050 km and for all inclination values, as shown in Figure \ref{fig:4}. The maneuver is performed when $a \approx 1.0497 \times 10^3$ km and $e \approx 2.03\times 10^{-4}$, in all cases.

\section{Conclusions}

Our results show that a long-term orbit has a semi-major axis with values between $810-100$ km and an eccentricity equal to $10^{-3}$. Furthermore, we show that for this eccentricity value the possible errors in Titania's gravitational coefficients can affect the probe's lifetime. We also show that specific values of ($\omega$) and ($\Omega$) contribute to increased probe life. In addition, we found a minimum value of $\Delta V$ to perform a maneuver capable of avoiding a collision of the probe with the surface of Titania and putting it back in a circular orbit. Our results may be extremely important for studies of sending a probe to this system.

\section{Acknowledgements}

The authors thank (CAPES) Financing Code 001. Project 2016/23542-1 from FAPESP. SMGW thanks CNPq (Proc 313043/2020-5) for the financial support. AA thanks the Center for Mathematical Sciences Applied to Industry (CeMEAI), funded by FAPESP (grant 2013/07375-0).

\bibliographystyle{unsrt}  


\end{document}